\documentclass[prb,twocolumn]{revtex4}

\usepackage{graphicx}
\usepackage{amsmath,amssymb}
\usepackage{amsbsy}
\usepackage{bm}

\begin{document}


\newcommand{\up}{\uparrow}
\newcommand{\down}{\downarrow}
\renewcommand{\d}{{\rm d}}
\newcommand{\e}{{\rm e}}
\newcommand{\imai}{{\rm i}}


\title{Coherent Transport through an interacting
double quantum dot: Beyond sequential tunneling}

\author{Jonas Nyvold Pedersen}
\author{Benny Lassen}\altaffiliation[Now at ]{Mads Clausen Institute,
University of Southern Denmark, Grundtvigs All{\'e} 150,
6400 S{\o}nderborg, Denmark.}
\author{Andreas Wacker}
\affiliation{Mathematical Physics, Physics Department,
Lund University, Box 118, 22100 Lund, Sweden}
\author{Matthias H. Hettler}
\affiliation{Forschungszentrum Karlsruhe, Institut f{\"u}r
  Nanotechnologie, Postfach 3640, 76021 Karlsruhe, Germany}

\date{\today}

\begin{abstract}
Various causes for negative differential conductance
in transport through an interacting double quantum dot are
investigated. Particular focus is given to the
interplay between the renormalization of the energy levels due to the
coupling to the leads and the decoherence of the states. The calculations
are performed within a basis of many-particle eigenstates and we
consider the dynamics given by the von Neumann-equation taking into
account also processes beyond sequential tunneling.
A systematic comparison between the levels of
approximation and also with different formalisms is performed.
It is found that the current is qualitatively well described by
sequential processes as long as the temperature is larger
than the level broadening induced by the contacts.
\end{abstract}
\pacs{73.23.Hk,73.63.Kv}

\maketitle

\section{Introduction}
The field of electronic transport through metallic and semi-conducting
quantum dots has been a topic of intense research for over a decade.
While in the early experiments the main focus was on charging effects
leading to the phenomenon known as `Coulomb
blockade',\cite{KastnerRMP1992}  and later on
the Kondo effect in single quantum dots,\cite{PustilnikJPC2004}
in recent years the attention
has shifted to more elaborate systems such as double quantum dot
systems.\cite{WielRMP2003,HatanoPRL2004,DicarloPRL2004,HuttelPRB2005,FasthNL2005,SigristPRL2006,BrandesPSSB2006}
In addition, the study of the electronic spectrum of quantum
dots (excited states) was possible as the physical size of quantum
dots could be further reduced by improved
lithographical methods as well as the rise of new materials such
as nanotubes and semi-conducting
nanowires.\cite{FuhrerNanoletter2007}

Already in the '90s, there were experiments\cite{WeisPRL1993}
on single quantum dots
displaying non-monotonous current-voltage characteristics,
for which the current decreases with increasing bias,
leading to a negative differential conductance (NDC).
Within the orthodox theory of sequential tunneling, such effects were
explained by the presence of excited states which were more weakly
coupled to the leads than the ground states (for a given charge on
the dot). The reason for a state dependent coupling could be either due to
spin (Clebsch-Gordon coefficients)\cite{WeinmannPRL1995} or the
different orbital
wave-functions of the various states.\cite{HettlerEPL2002,RoggeNJP2006}
A bias voltage dependence of the lead-dot tunnel coupling can
lead to a weak NDC effect, see, e.g., Ref.~\onlinecite{NauenPRB2004}.

For a double quantum dot (DQD) system  with the two quantum dots
in series, very sharp current peaks (and corresponding NDC) was
observed in experiments where the inter-dot coupling $\Omega$ was
weak.\cite{WielRMP2003} Depending on whether two levels on
different dots are `aligned' or not, the current will be high
(in alignment) or low (off alignment). As the quantum levels in
the different dots are  shifted differently by the applied bias
voltage (depending on the various capacitances of the system) the
alignment condition is fulfilled at certain small ranges of the
bias voltage, leading to the current peaks.

In this paper we focus on quantum transport through a DQD.
This system  offers the possibility to
study the interplay between the coherent quantum mechanical
oscillation inside the DQD and the influence of the coupling to
leads. Especially, we are interested in the negative differential
conductance caused by this interplay (rather than the well-known
sources discussed above).
In the regime where the inter-dot tunneling coupling dominates
over the coupling to leads, it was found in
Ref.~\onlinecite{AghassiPRB2006} that NDC only occurs if the
spatial symmetry of the system is broken, e.g. due to asymmetric
coupling to leads or detuning of the bare level
energies.\footnote{The results in Ref.~\onlinecite{AghassiPRB2006}
were derived for a double dot including spin, but the conclusion
can be transferred to the spin-less case.}
In another recent paper,
B. Wunsch {\it et al.} [\onlinecite{WunschPRB2005}] investigated
the transport in the opposite limit, where the inter-dot coupling
is weak and with a small detuning of the bare level energies, i.e.
for an asymmetric system. They found that NDC can be caused by a
level renormalization due to the coupling to leads. Also in the
weak inter-dot coupling regime, Djuric {\it et al.}
[\onlinecite{DjuricJAP2006}] found NDC even for symmetric systems
for certain ratios between the inter-dot tunneling coupling and the
coupling to leads. The effect was explained in terms of
decoherence due to the coupling to leads, which depended
on the occupation of the dot. In this article we show how these
effects relate to each other.

In the above mentioned works, the current was only calculated to
lowest (first) order in the lead-dot tunnel coupling, so strictly
speaking, the results are only valid in the sequential tunneling
limit. However, for the issues addressed above,
the coupling to the contacts strongly influences not only the
occupations but also affects the nature of the transport,
  especially its quantum-mechanical coherence.
Thus it is not a priori clear, if the first-order approach is
appropriate, even if the temperature is higher than the energy scale
(line-width) due to the coupling to the leads.
Below, we investigate
the current to higher order in the lead-dot coupling by applying
the method described in Ref.~\onlinecite{PedersenPRB2005a} and
compare with first order results. For both cases we find
qualitative agreement if the temperature is higher or comparable to the
line-width due to coupling to the leads, but we also discuss the
behavior for lower temperatures, where the first order approach
becomes unreliable.

The paper is organized as follows: in section
\ref{sec:model} we present in detail the DQD model system we consider.
Section \ref{sec:approach} discusses in brief the transport models
we are using to obtain the transport results
(details are presented in the appendix). The
potential sources of NDC behavior are discussed in detail in section
\ref{sec:ndc_sources}. Finally we summarize
our findings in section \ref{sec:Summary}.

\section{The model system}
\label{sec:model}
We consider a double quantum dot system, where the spin
degree of freedom has been omitted in order to simplify the
analysis (in subsection \ref{sec:doublespin} the double spin case is
briefly addressed). In a single-particle basis the Hamiltonian for the
system reads
\begin{multline}
H=E_\alpha d_\alpha^\dag d_\alpha^{{}} +E_\beta d_\beta^\dag
d_\beta^{{}}
+U d_\alpha^\dag d_\alpha^{{}} d_\beta^\dag d_\beta^{{}}
+\left(\Omega d_\beta^\dag d_\alpha^{{}}
+h.c\right)\\
+\sum_{k\ell}E_{k\ell}^{{}} c^\dagger_{k\ell} c^{{}}_{k\ell}
+\sum_{k}\Big( t_{kL}^{{}}d_\alpha^\dag c_{kL}^{{}}
+t_{kR}^{{}}d_\beta^{{\dag}}c_{kR}^{{}}+h.c.\Big),
\label{EqHamiltonDQD}
\end{multline}
where the first line describes the isolated quantum dot system
with $U$ being the Coulomb energy for occupying both dots,
$\Omega$ the inter-dot tunneling coupling and with $\alpha/\beta$
denoting the left/right dot. The first term in the second line
accounts for the leads with index $\ell=L/R$ for the left/right lead
and levels counted by $k$.
The  last term is the lead-dot tunneling coupling.
We parameterize the lead-dot coupling parameters $t_{k\ell}$ by
$\Gamma_\ell(E)=2\pi\sum_{k}|t_{k\ell}|^2\delta(E-E_{k\ell})$.
Here we use the constant value $\Gamma_\ell$ for $|E|\leq 0.95W$ and
assume  $\Gamma_\ell(E)=0$ for $|E|>W$. For $0.95W<|E|<W$ we
interpolate with an elliptic behavior in order to avoid
discontinuities. Furthermore, we define
  $\Gamma=\Gamma_L+\Gamma_R$.
The bias voltage $V_\mathrm{bias}$ is applied symmetrically to the
  electrochemical potentials of both leads,
$\mu_L=-\mu_R=eV_\mathrm{bias}/2$, where $e$ is the positive elementary charge.

Throughout this paper we include the Coulomb interaction
by considering a basis of
many-particle states for the isolated DQD, which allows for a
consistent description of many-particle effects, see also
Ref.~\onlinecite{MuralidharanPRB2006} and references given therein.
Thus, we diagonalize the first line
of the Hamiltonian from Eq.~(\ref{EqHamiltonDQD}) and find
the eigenstates and the corresponding energies. $E_0=0$ is the
energy of the empty state,
$E_1=\rho-\sqrt{\Delta^2+4\Omega^2}/2$ and
$E_2=\rho+\sqrt{\Delta^2+4\Omega^2}/2$ the
energies of the single occupied states, and
$E_d=E_\alpha+E_\beta+U$ the energy of the double occupied state,
where $\Delta=E_\alpha-E_\beta$ and $\rho=(E_\alpha+E_\beta)/2$. The
states with energies $E_1/E_2$ are referred to as the
bonding/anti-bonding state.

Depending on the occupation of the dot states different transport
regimes can be defined. Current through the DQD is effectively blocked if
no one-particle excitation lies in the bias window
between the Fermi levels of both contacts. This is known as the Coulomb
blockade regime. Therefore, as the bias is increased, a current
can  flow through the structure
whenever a one-particle excitation becomes
energetically allowed, leading to a step feature in the current and a
corresponding peak in the differential conductance. Having 4 such possible
excitations ($0\leftrightarrow 1,2$, and  $1,2\leftrightarrow d$),
at most four steps can be observed in the $IV$-curve.
Further steps can be seen, if spin is considered
as well.\cite{BulkaPRB2004}

The quantum rate equations from
Refs.~\onlinecite{GurvitzPRB1996,StoofPRB1996,ElatariPLA2002} are
valid in the high-bias limit, i.e. if the energy difference between
the chemical potentials in the contacts and the excitations exceeds
both the level broadenings $\Gamma_{\ell}$ and the temperature.
If only one-particle states are within the bias window,
but double occupation is forbidden, i.e.,
$(E_d-E_{1}),(E_d-E_{2})\gg \mu_L,\mu_R$,
these equations provide the plateau current
\begin{equation}\label{EqStoof}
I_1=\frac{e}{h} \frac{\Omega^2\Gamma_R}{\Omega^2(2+\Gamma_R/\Gamma_L)+\left(\Gamma_R/2\right)^2+\Delta^2}\, .
\end{equation}
If, in contrast, all excitations are within the bias window, one
obtains
\begin{equation}\label{EqElatari}
I_2=\frac{e}{h} \frac{\Gamma_L\Gamma_R\Gamma\Omega^2}{(4\Omega^2+\Gamma_L\Gamma_R)(\Gamma/2)^2
+\Delta^2\Gamma_L\Gamma_R},
\end{equation}
These values will be compared to our calculations in the subsequent
sections.

\section{The von Neumann approach}
\label{sec:approach}
Our calculations are based on the von Neumann equation for the
density matrix, as
described in detail in Ref.~\onlinecite{PedersenPRB2005a}.
The key idea is to use a set of many-particle states labeled
$|a\rangle,|b\rangle,\ldots$, with energies $E_a,E_b,\ldots$,
respectively, which diagonalize the Hamiltonian of the
system, including the many-particle interaction. (In our case these
are the states $|0\rangle,|1\rangle,|2\rangle,|d\rangle$ introduced in
section \ref{sec:model}.) Transport occurs by tunneling of electrons
with a quantum number $k$ from a lead $\ell$ into the system, while
the state is changed from an $N$-particle state $|a\rangle$ to an
$N+1$-particle state $|b\rangle$. The corresponding matrix element
is $T_{ba}(k\ell)$.
In Ref.~\onlinecite{PedersenPRB2005a} the full correlations of up to
two particles entering and leaving the system was taken into account
and this
approach will be referred to as the {\em second order von Neumann approach}
(2vN) in the following. In addition, we apply the same concept
restricting to single electron processes, which we call the {\em first
order von Neumann approach} (1vN). The resulting equations for the 1vN
approach are given in App.~\ref{AppFirstOrder}.
Both the first and the second order approaches include
the non-diagonal elements of the density matrix,
which allows us to consider the regime of
both weak and strong inter-dot coupling $\Omega$.
This is demonstrated in  App.~\ref{AppComparison}, where
our method is also compared with other approaches.
The 1vN approach
neglects level broadening effects  (of order $\Gamma$)
and is thus expected to be valid only  for
$k_BT\gg \Gamma$ or in the high-bias limit. In contrast, the
2vN approach is able to reproduce such effects and gives good results
above the Kondo temperature.\cite{PedersenPRB2005a}
Due to the self-consistency, the 2vN approach contains in addition
to two-particle correlations also a subset of higher order correlations.
In App.~\ref{AppAnderson} we show that the result for the Anderson
model is identical with the corresponding result obtained from
the Real-Time diagrammatic approach
in resonant tunneling approximation.\cite{KonigPRL1996,KonigPRB1996}

The 1vN approach contains sums of the form
$\sum_{k}T_{ba}T^*_{b'a'}/(E_k-E_{b'}+E_a-\imai 0^+)$  (see
App.~\ref{AppFirstOrder}). Decomposing
\[\begin{split}
\frac{1}{E_k-E_{b'}+E_a-\imai 0^+}={\cal P}
\left\{\frac{1}{E_k-E_{b'}+E_a}\right\}\\
-\imai \pi  \delta(E_k-E_{b'}+E_a)\, ,
\end{split}\]
the imaginary part can be related to electronic transition rates, while the
real part acts as an effective renormalization of the
transition energies between different many-particle states.
In some calculations we will neglect all terms resulting from
these real parts (we denote this by ``no real parts'') in
order to demonstrate their relevance.

\section{Sources for NDC behavior}
\label{sec:ndc_sources} 
\begin{figure}
  \includegraphics[width=0.45\textwidth]{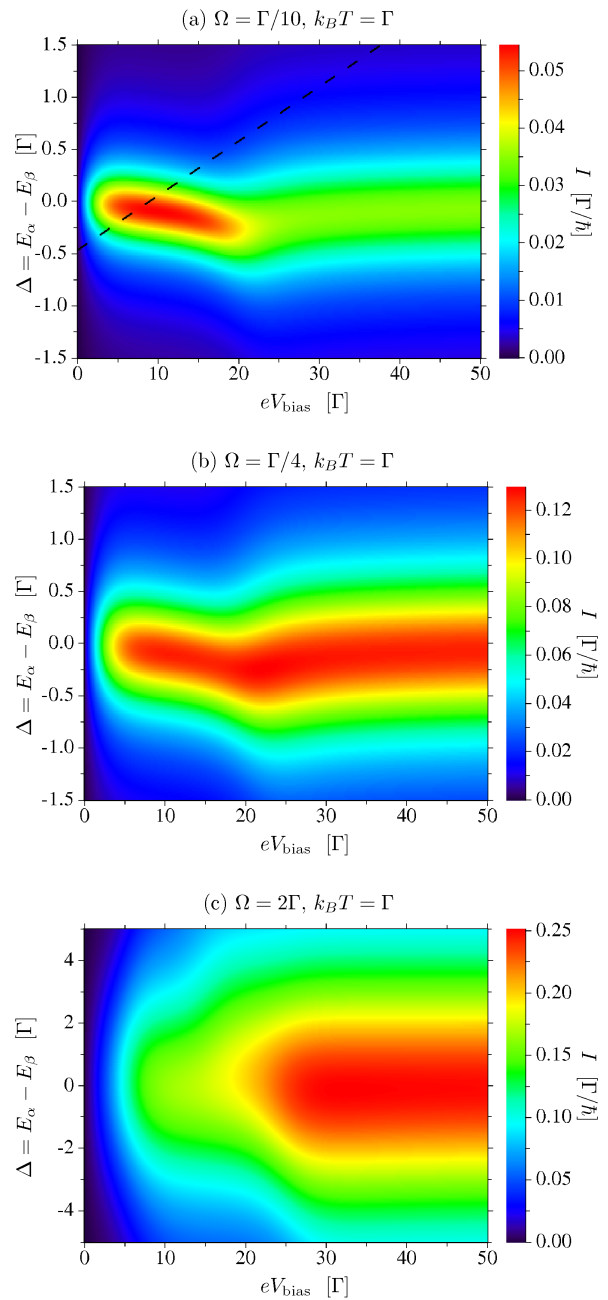}
  \caption{Current versus  bias voltage and detuning
    $\Delta=E_\alpha-E_\beta$ for
    (a) $\Omega=\Gamma/10$,
    (b) $\Omega=\Gamma/4$, and
    (c) $\Omega=2\Gamma$ using the first order approach (1vN) including
        the real parts. The other parameters are $E_\alpha+E_\beta=0$,
    $\Gamma_L=\Gamma_R=\Gamma/2$, $k_BT=\Gamma$, $U=10\Gamma$ and
    $W=35\Gamma$.}
\label{FigCondPlots}
\end{figure}

In a real experimental double-dot
structure, the applied source-drain bias $V_\mathrm{bias}$ does
not only determine the electrochemical potentials in  the leads,
but it  will also shift the dot level energies by polarization.
The amount of these shifts depends on the details of
the various dot capacitances and can be taken into account by
lever arm factors $\lambda_{\alpha}$, $\lambda_{\beta}$ for the
respective dot levels\cite{WielRMP2003}. In addition, if gates
are present, the gate voltages
$V^{\alpha/\beta}_{\mathrm{gate}}$ can  also shift the respective
dot levels with efficiency factors $\eta_{\alpha},\eta_{\beta}$.
Therefore, the
voltage dependence of the dot level energies can be written as
\begin{eqnarray}
E_\alpha &=& E_{\alpha}^0+ \lambda_\alpha \frac{e
V_\mathrm{bias}}{2}-\eta_{\alpha} eV^{\alpha}_{\mathrm{gate}},\\
E_\beta &=& E_{\beta}^0-\lambda_\beta \frac{e
V_\mathrm{bias}}{2}- \eta_{\beta} eV^{\beta}_{\mathrm{gate}}
\end{eqnarray}
with $E_i^0$ being the equilibrium level of the energies.
This allows for an independent control of $V_\mathrm{bias}$,
the level difference (detuning) $\Delta=E_\alpha-E_\beta$,
and the average level
$\rho=(E_\alpha+E_\beta)/2$. In the following we set $\rho=0$, meaning that
the dot states are at equal energetic distance from the equilibrium Fermi
level. In Fig.~\ref{FigCondPlots} we show the current
calculated with the 1vN approach as a function of $V_\mathrm{bias}$ and $\Delta$ for different
inter-dot coupling strengths $\Omega$.

In a real experiment, the current voltage characteristic
corresponds typically to a line in the
$(V_\mathrm{bias},\Delta)$-plane. With zero gate voltages, one has
\[
\Delta=E_\alpha^0-E_\beta^0+
(\lambda_\alpha+\lambda_{\beta})\frac{e V_\mathrm{bias}}{2}\, ,
\]
so one follows a straight line with positive slope, see, e.g., the
dashed line in Fig.~\ref{FigCondPlots}(a). For a
sufficiently large slope $(\lambda_\alpha+\lambda_{\beta})/2$,
we observe first an increase of
current for low bias, and then a decrease of the current
as the levels move out of resonance with increasing $\Delta$.
This is the standard NDC effect induced by electrostatic polarization.

\begin{figure}
  \includegraphics[width=0.48\textwidth]{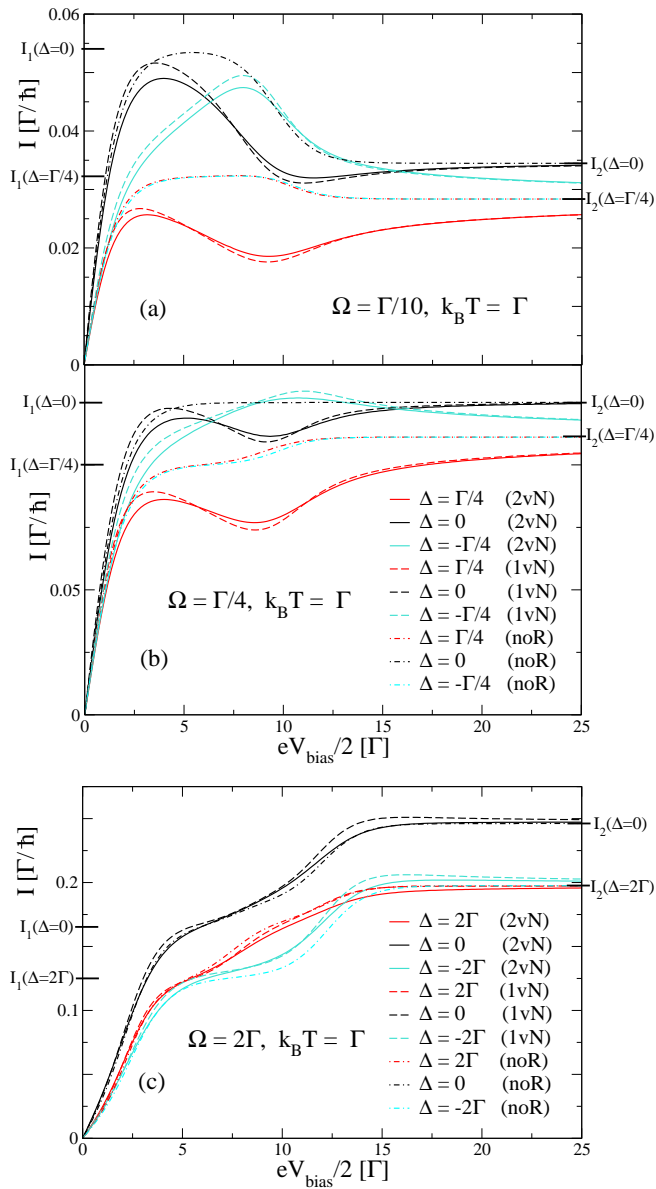}\\
  \caption[a]{
  Current versus bias voltage for
  different detunings $\Delta$. All other values like in
  Fig.~\ref{FigCondPlots}. The dashed lines are the 1vN results
  including the real parts, the dashed-dotted lines are the 1vN
  results without the real parts (noR), and the full lines are the 2vN
  results.  The values from Eqs.~(\ref{EqStoof},\ref{EqElatari}) for
  $\Delta=0$ and $\Delta=\pm\Gamma/4$ or $\Delta=\pm2\Gamma$
  (which do not depend on the sign of $\Delta$) are shown
  on the $y$-axis. We only show positive bias here,
  as the negative bias result corresponds to the results with the
  opposite sign of $\Delta$ for the symmetric coupling to contacts
  considered here.}
\label{FigCurveAll}
\end{figure}

In addition to the above `trivial' effect
we can identify two further scenarios for NDC for a fixed detuning
$\Delta$\footnote{A constant $\Delta$  can be realized by
minimizing the polarization of the dot levels from the leads
(vanishing $\lambda$) or by compensating with appropriate
gate voltages proportional to $V_\mathrm{bias}$.}, which
will be discussed in detail below.
Firstly, we notice that for small and intermediate $\Omega$
[Figs.~\ref{FigCondPlots}(a,b)]
the current peak (red region)
is shifted to negative $\Delta$ with increasing bias
$0<eV_\mathrm{bias}<20 \Gamma=2U$.
Thus, the current drops with $V_\mathrm{bias}$
if $\Delta\gtrsim 0$ is kept constant,
see also the dashed lines in Fig.~\ref{FigCurveAll}(a,b).
A more detailed analysis is given in Sec.~\ref{SecDetuning}.

Secondly, the height of the
current peak (at fixed bias) drops when the bias voltage
exceeds $2U=20 \Gamma$ in Fig~\ref{FigCondPlots}(a) where the
inter-dot tunneling coupling is much weaker than the coupling to
leads.  This provides NDC around $\Delta\approx 0$ as discussed in detail in
Sec.~\ref{SecDecoherence}. With increasing inter-dot coupling this
effect vanishes as seen in Figs.~\ref{FigCondPlots}(b-c).

\subsection{First order versus second order}
\begin{figure}
  \includegraphics[width=0.475\textwidth]{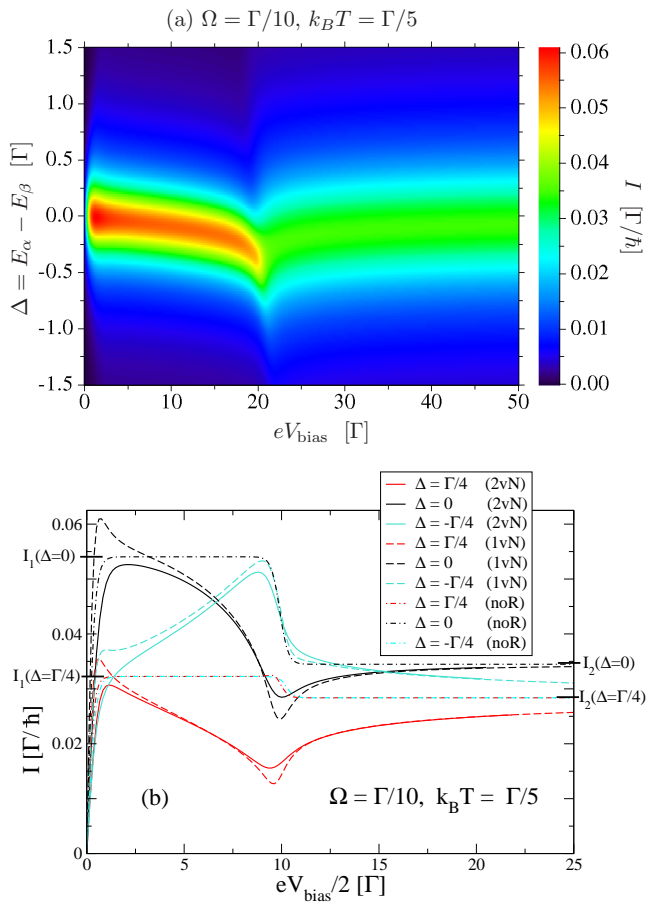}\\
  \caption[a]{Results for low temperature $k_BT =\Gamma/5$ and
$\Omega= \Gamma/10$. All other parameters as
Figs.~\ref{FigCondPlots},\ref{FigCurveAll}.
Upper panel: Results of the 1vN approach. Lower panel: 
Comparison of 1vN, 2vN and the 1vN without real parts (noR) approaches. 
The 1vN  strongly exaggerates the features at the current steps for
this low temperature in comparison to the 2vN approach. The 1vN
without real parts is qualitatively incorrect.}
\label{FigLowT}
\end{figure}

In Fig.~\ref{FigCurveAll} we provide a systematic comparison
between the 1vN approach (dashed line) and the 2vN approach (full
line) for different values of the detuning $\Delta$, corresponding
to cuts along horizontal lines in Fig.~\ref{FigCondPlots}.
We find that both approaches are in good qualitative agreement
both for large and small values of the interdot coupling $\Omega$ at
the considered temperature. This shows that the 1vN approach works
well even for the moderate temperature $k_BT=\Gamma_L+\Gamma_R$.
We observe small discrepancies close to current steps, where
the broadening is underestimated due to the neglect of line-width
broadening in the 1vN approach.
As expected, these discrepancies are strongly enhanced if the temperature
drops below the level broadening as shown in Fig.~\ref{FigLowT}(b) for
$k_BT =(\Gamma_L+\Gamma_R)/5$.

\subsection{NDC due to level renormalization}\label{SecDetuning}

Let us now focus towards the bias range $eV_\mathrm{bias}<2U$, where
the double occupied state does not yet contribute to the current.
In this regime we observe a significant shift
of the current peak from its naively expected position at
$\Delta=0$ for small and intermediate $\Omega$
[Figs.~\ref{FigCondPlots}(a,b)].

In Ref.~\onlinecite{WunschPRB2005} Wunsch {\it et al.}
considered a DQD including spin. They calculated
the transport using a first-order diagrammatic real-time transport
approach (see e.g.
Refs.~\onlinecite{KonigPRL1996,KonigPRB1996}), restricting
themselves to the limit of small inter-dot coupling $\Omega\ll
\Gamma$, where electronic states localized on the single dots form
an appropriate basis.
For positive $\Delta$ they observe pronounced NDC
similar to Fig.~\ref{FigCurveAll}(a,b). This was explained in the
following way: The energy levels $E_\alpha,E_\beta$ are renormalized
due to the couplings to the contacts. For finite $U$ this renormalization
is strongest if the levels are close to the chemical potential. Now,
the localized states used in Ref.~\onlinecite{WunschPRB2005}
couple mostly to the nearest lead and thus the renormalization
differs for both levels at finite bias. This provides a
bias-dependent renormalized $\Delta_{\mathrm{eff}}$ and the
maximum of the current occurs at $\Delta_{\mathrm{eff}}=0$ rather
than $\Delta=0$.
This effect does not occur for $U=0$
because in this case the renormalization does not depend on the
location of the chemical potential (this becomes obvious in a
Green function treatment providing the exact result for $U=0$).

Our results in Figs.~\ref{FigCondPlots}(a,b) are in full agreement
with these findings. In particular, the shift of the current peak
position can be directly attributed to a
bias-dependent renormalization $\Delta\to \Delta_{\mathrm{eff}}$,
where $\Delta_{\mathrm{eff}}>\Delta$ for positive bias.
This  renormalization is also reflected in the magnitude of the current:
For $eV_\mathrm{bias}<2U$,  Eq.~(\ref{EqStoof}) suggests the
plateau current $I_1$. In contrast, no such plateau is seen in
Fig.~\ref{FigCurveAll}(a,b) for small and
intermediate tunnel coupling. Furthermore, the calculated
currents are smaller than the respective value of $I_1(|\Delta|)$
for $\Delta=0$ and
$\Delta=\Gamma/4$, while they exceed $I_1(|\Delta|)$ for
$\Delta=-\Gamma/4$. This can be fully attributed to a bias-dependent
renormalized $\Delta_{\mathrm{eff}}>\Delta$ which should be used in
Eq.~(\ref{EqStoof}).

Fig.~\ref{FigCondPlots}(c) shows that the effect
vanishes for strong inter-dot coupling $\Omega> \Gamma$, a regime
where the approach of Ref.~\onlinecite{WunschPRB2005} fails.
In addition, the quantitative agreement between the 1vN and 2vN
approach shows that the shift of the resonance condition is not
an artefact of a first order tunneling approach (such as 1vN or the
approach used in Ref.~\onlinecite{WunschPRB2005}) but persist even
if higher order tunneling processes are taken into account, which,
e.g., cause line-width broadening.

\subsection{Symmetry with respect to the sign of $\Delta$}

The quantum rate equation results 
Eqs.~(\ref{EqStoof},\ref{EqElatari}) 
only depend on the absolute value $|\Delta|$. In contrast,
Figs.~\ref{FigCondPlots},\ref{FigCurveAll} exhibit a strong
asymmetry with respect to the sign of $\Delta$.
For the cases $\Omega < \Gamma$ (Fig.~\ref{FigCurveAll} 
(a,b) ) the reason lies mostly in the level renormalization discussed 
above. This can be deduced from the fact that the curves without real
parts (noR) that neglect the level renormalization are actually approximately
symmetric. 
However, for $\Omega> \Gamma$ [Fig.~\ref{FigCurveAll} (c)]
all numerical approaches give an additional step at 
$eV_{\rm bias}/2\approx E_1+U$ for positive $\Delta = 2 \Gamma$, 
which is absent for negative  $\Delta$. 

For the case of large $\Omega$ 
the ``molecular'' bonding and anti-bonding states provide an 
appropriate description of the single-electron states of the DQD
(see also the comparison in App.~\ref{AppComparison}).
Because of the large (positive) detuning $\Delta > \Gamma, k_B T$ the wave
function of the bonding state has much more weight on the right dot
(interface to the collector lead) than on the left dot. 
Correspondlingly,  the anti-bonding state has more weight on the left
dot (interface to the emitter electrode). This asymmetry in the wave
functions leads to an effective asymmetry of coupling of these states
to the leads, e.g. for positive $\Delta$, the anti-bonding
state is strongly coupled to the emitter, whereas the bonding state is 
strongly coupled to the collector lead.

The eigenenergies $E_{1,2}$ of the bonding and anti-bonding state 
are given by $\mp \sqrt{\Delta^2+4\Omega^2}/2 = \mp \sqrt{5} \Gamma$ (see
section~\ref{sec:model}). As $E_1 = - \sqrt{5} \Gamma$ is the only
state with negative energy, it is the overall ground state of the DQD.
This means that in equilibrium (zero bias) the bonding state is mainly occupied,
whereas all other states have only small occupation probability. 
The first current step appears at a bias $eV_{\rm bias}/2 \sim  |E_1|
\approx 2.2 \Gamma$ 
when the DQD can be emptied. At the same bias the anti-bonding state
can be occupied by tunneling of an electron from the emitter into the 
empty DQD. Because of the coupling asymmetry, above a bias 
$eV_{\rm bias}/2 \sim  |E_1|$ there is a large change in the average
occupation of the DQD states: the bonding state is strongly depleted
due to the good coupling to the collector lead, whereas the
anti-bonding state is now favorably filled by electrons tunneling into
the DQD from the emitter lead. Thus, on the first plateau the DQD is
most of the time in the anti-bonding state.

As the bias is further increased, the doubly
occupied state comes within energetic range when the bonding state can
also be filled {\it in addition} to the anti-bonding state. This happens
at   $eV_{\rm bias}/2 = E_1+U = (-\sqrt{5} + U ) \Gamma \approx 7.8
\Gamma$, which is where the second current step sets in. Now, the anti-bonding
state loses occupation in favor of the double occupied state and the
bonding state. Finally, at even larger bias $eV_{\rm bias}/2 = E_{2}+U =
(\sqrt{5} +10) \Gamma \approx 12.2 \Gamma$, the doubly occupied state
can be populated by an electron tunneling in from the emitter,
even if the DQD is previously in the bonding state. This leads to the (weak)
third current step for the curves corresponding of positive values of 
bias and $\Delta$.  

In contrast, for negative $\Delta$ and positive bias, because of the reversed
spatial asymmetry of bonding and anti-bonding wave functions, the
DQD remains mostly in the bonding state, even though the bias has
exceeded  $eV_{\rm bias}/2 \sim  |E_1| \approx 2.2 \Gamma$. The
occupation of the anti-bonding state remains very small in the range 
$ 2.2 \Gamma < eV_{\rm bias}/2 < 12.2 \Gamma$. Therefore, the 
middle current plateau is strongly suppressed and smeared out by the thermal 
(and linewidth) broadening. 

The quantum rate equation results Eqs.~(\ref{EqStoof},\ref{EqElatari}) 
correspond to the current values of the first and third current
plateau. To capture the middle plateau one would have to account for
the possible transitions from the anti-bonding state to the doubly
occupied state in the relevant bias range. This was not done in
Ref. \onlinecite{DjuricJAP2006}, consequently the middle plateau
is never observed in their Fig. 6  even for the case $\Omega = \Gamma$.

\subsection{NDC due to decoherence}\label{SecDecoherence}
Now we focus on the behavior around $eV_\mathrm{bias}\approx 2U$, where
the double occupied state enters the window between the left and the
right chemical potential. Fig.~\ref{FigCondPlots}(a) shows that
the current peak is significantly larger for
$eV_\mathrm{bias}< 2U$ (region 1) than for $eV_\mathrm{bias}> 2U$
(region 2) for small $\Omega$. Just the opposite holds for
large $\Omega$, see Fig.~\ref{FigCondPlots}(c).

This drop of current has been addressed by Djuric {\it et al.}
\cite{DjuricJAP2006}, using the quantum
rate equation formalism developed in
Ref.~\onlinecite{DongPRB2004}. They restricted the analysis to
a fully symmetric system, i.e. $\Delta=0$, and did not include any
real parts.
In this case Eqs.~(\ref{EqStoof},\ref{EqElatari}) provide
$I_1>I_2$ for $\Omega< \sqrt{\Gamma_L\Gamma_R}/2$, and
$I_1<I_2$ for $\Omega> \sqrt{\Gamma_L\Gamma_R}/2$.
Numerically, they observe a smooth interpolation between these
plateau values upon variation of $V_\mathrm{bias}$,
similar to the result of
the 1vN approach without real parts (dot-dashed line) in
Fig.~\ref{FigCurveAll}. This gives rise to
NDC around  $eV_\mathrm{bias}\approx 2U$
for weak inter-dot coupling $\Omega< \sqrt{\Gamma_L\Gamma_R}/2$.

While the observed increase of current for
$\Omega> \sqrt{\Gamma_L\Gamma_R}/2$ can be easily attributed to the
opening of a new current channel, the NDC for
$\Omega< \sqrt{\Gamma_L\Gamma_R}/2$ is less straightforward.
In the limit of small $\Omega$, Eqs.~(\ref{EqStoof},\ref{EqElatari})
read
\[
I_1\approx \frac{e}{h} \Omega^2
L(\Delta,\Gamma_R)\, ,
\qquad
I_2\approx \frac{e}{h} \Omega^2
L(\Delta,\Gamma_R+\Gamma_L)
\]
with the Lorentzian $L(\Delta,\gamma)=\frac{\gamma}
{\Delta^2+\left(\gamma/2\right)^2}$.
This is just the expression from Fermi's
golden rule for sequential tunneling between the localized dot
states, which is limiting the current for weak coupling, see
Ref.~\onlinecite{SprekelerPRB2004}. The broadening $\gamma$ is given by the
dephasing of the coherence between the localized dot states
\cite{KiesslichPRB2006}. For the second plateau $I_2$
the inter-dot Coulomb repulsion
does not play a role and we get the broadening  $\gamma= \Gamma_R+\Gamma_L$.
However, for  $eV_\mathrm{bias}< 2U$, the left contact cannot add
an electron to the system if one electron is already present in
the DQD. Thus, the dephasing of the coherent
transitions between the dots is only due to
$\gamma=\Gamma_R$. Therefore, in Fig.~\ref{FigCondPlots}(a),
the current peak (at fixed bias voltage)
is higher and narrower (as a function of $\Delta$)
for  $eV_\mathrm{bias}< 2U$ (first plateau) in comparison to
$eV_\mathrm{bias}> 2U$ (second plateau).

As displayed by Fig.~\ref{FigCurveAll} the transition between the
plateau values for $\Delta=0$ is much more complex than the smooth
transition suggested by Ref.~\onlinecite{DjuricJAP2006}. Both the
1vN and the 2vN approach do not reach a stable value at the first
plateau even for $U$ being by far the largest energy. Instead the
current drops over the full length of the plateau, giving a much
weaker NDC than predicted by Ref.~\onlinecite{DjuricJAP2006}. The
plateau value $I_2$ is indeed reached in region 2 for
$V_\mathrm{bias}\to \infty$, but due to the level renormalization
effect the transition is much broader than if it was only given by
temperature (or even a combination of temperature and line-width
broadening $\Gamma$). For $\Omega=\sqrt{\Gamma_L\Gamma_R}/2$
(Fig.~\ref{FigCurveAll}(b)) the current does not reach the plateau
value $I_1$, and at the transition between the two regimes
($eV_\mathrm{bias}\simeq 2U=20\Gamma$) a dip in the current is
observed. For $\Omega=2\Gamma$  (see Fig.~\ref{FigCurveAll}(c))
where $I_1<I_2$ and no NDC should occur due to decoherence, we
observe that the plateau values are indeed reached in both
regimes. While Fig.~\ref{FigCurveAll}(a) shows very non-monotonous
behavior, Fig.~\ref{FigCondPlots} shows that the current varies
more continously in the  $(\Delta,V_\mathrm{bias})$ plane. This
shows that the main difference between our 1vN approach  and the
result of Ref.~\onlinecite{DjuricJAP2006} is the level
renormalization discussed above resulting in an effective
$\Delta_{\mathrm{eff}}$, which is not contained in the quantum
rate equation formalism of Ref.~\onlinecite{DongPRB2004}.

\begin{figure}
  \includegraphics[width=0.48\textwidth]{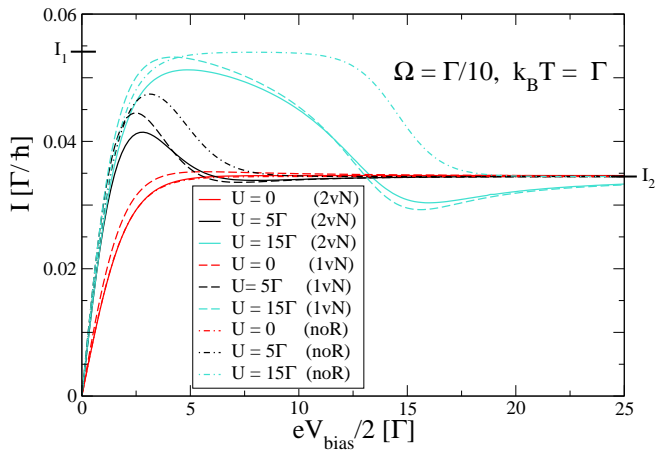}\\
  \caption{Results for $\Delta=0$, $\Omega=\Gamma/10$, and
   different values of $U$.  All other parameters as in
   Figs.~\ref{FigCondPlots},\ref{FigCurveAll}.}
   \label{FigCoulomb}
\end{figure}

So far we have only considered a fixed value of the Coulomb
repulsion $U=10\Gamma$. Fig.~\ref{FigCoulomb} shows results
for different values of $U$ at weak coupling $\Omega=\Gamma/10$ and
zero detuning $\Delta=0$. It should be noted that even for $U$
being by far the dominant energy (e.g. $U=15\Gamma$) no clear plateau
value is observed, but instead a slow cross-over between the two
regimes. Again, relatively small differences between the 1vN and 2vN
approach are observed, though in general the 2vN approach
reduces somewhat the level renormalization effect and leads to
more plateau-like current-bias characteristic
than the 1vN approach. Here, a particular
surprising feature is the fact that a small (unphysical) NDC
remains even for $U=0$ in the 1vN approach (see also the
analytical result in App.~\ref{AppComparison}), which is however
absent in the 2vN approach (in full agreement with the
transmission result by Green functions).

We conclude that the NDC-scenario due to decoherence outlined
in Ref.~\onlinecite{DjuricJAP2006} is strongly modified
by the level renormalizations addressed in section \ref{SecDetuning}.
Also note that the high-bias limit
$I_2$ is only reached for very high bias in the case of weak or intermediate
inter-dot coupling $\Omega$ as shown in  Fig.~\ref{FigCurveAll}.

\subsection{Transport with both spins} \label{sec:doublespin}
\begin{figure}
  \includegraphics[width=0.5\textwidth]{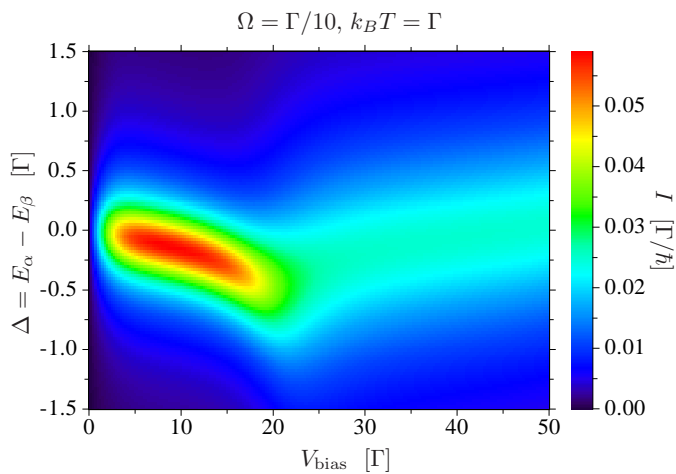}\\
  \caption{The current taking into account both spin
  directions and an
  intradot Coulomb repulsion of  $U_{\mathrm{intra}}=30\Gamma$
  calculated by the 1vN approach. All other
  parameters as Fig.~\ref{FigCondPlots}(a).}
\label{FigDoublespin}
\end{figure}

While we restricted ourselves to the case of spinless
 fermions before, we now take
into account both spin directions in both dots, thereby accounting
for sixteen different many-particle states. We add an intradot
Coulomb repulsion of $U_{\mathrm{intra}}=30\Gamma$ between the
different spin states within each dot, while the interdot Coulomb
repulsion $U=10\Gamma$ is assumed to be independent on the spin
direction. The result from the 1vN approach is shown in
Fig.~\ref{FigDoublespin}, which is qualitatively similar to
Fig.~\ref{FigCondPlots}(a) where spinless electrons were
considered. However, quantitatively, the features of level
renormalization and NDC due to coherence are enhanced.

\section{Discussion and summary}\label{sec:Summary}
We have studied transport trough a double quantum dot system and
investigated various sources of negative differential conductance (NDC)
potentially observable in experiment. Our method reproduces
the basic features of
Refs.~\onlinecite{WunschPRB2005,DjuricJAP2006}, which constitute
certain limits of our full numerical approach. In particular we
can treat all values of inter-dot coupling and
detuning of dot levels within the 1vN approach.
Further effects of line-width
broadening due to higher order tunneling events can be taken into account by
the 2vN approach. The NDC due to
level renormalization (Sec.~\ref{SecDetuning}) as introduced by
Ref.~\onlinecite{WunschPRB2005} using a strictly sequential
tunneling scenario is only quantitatively modified by the higher
order tunneling events (even at temperatures comparable to the
line-width broadening), thus showing a surprising robustness. The
NDC due to decoherence\cite{DjuricJAP2006}
(Sec. \ref{SecDecoherence}) is clearly
seen in the heights of the current peaks in the bias-detuning plane, see
Fig.~\ref{FigCondPlots}. Nevertheless this
effect is strongly masked by the level renormalization effects
if a constant detuning $\Delta$ is considered. Both effects show the
relevance of a consistent treatment of first-order tunneling terms,
which can be achieved by the 1vN approach discussed here.
Comparison with the higher-order 2vN approach\cite{PedersenPRB2005a}
for a large variety of parameters shows
that the validity of these first-order approaches is only
restricted to the temperature being larger than the level broadening.

\acknowledgments
This work was supported by the Swedish Research
Council (VR) and  the Villum Kann Rasmussen fond.

\appendix
\section{First order von Neumann approach}\label{AppFirstOrder}

We denote the many particle states by $|a\rangle$, $|b\rangle
\ldots$ with energies $E_a,E_b,\ldots$, respectively, and we use
the convention, that the particle number follows the position of
the letter in the alphabet, i.e. $|a\rangle$ is an $N$-particle
state and $|b\rangle$ is an $N+1$-particle state. The tunnel
matrix elements $T_{ba}(k\sigma\ell)$ for a transition from the
state $|a\rangle$ to $|b\rangle$ by entering of an electron from the
contact $\ell\in \{L,R\}$ with spin $\sigma$ and momentum $k$ can
be directly related to the single particle tunneling Hamiltonian
as given in Eq.~(\ref{EqHamiltonDQD}), see App. A of
Ref.~\onlinecite{PedersenPRB2005a}. The key quantities are the
elements of the reduced density matrix
$w_{b'b}=\text{Trace}\left\{|b\rangle \langle
b'|\hat{\rho}\right\}$, where the diagonal elements are the
probabilities to find the respective many-particle state and the
off-diagonal elements refer to correlations between the
many-body states induced via coherent tunneling processes to and from
the leads.

The equation of motion for $w_{b'b}$ is derived from the
von Neumann equation for the density matrix, see Eq.~(11) of
Ref.~\onlinecite{PedersenPRB2005a}, and depends on the
current amplitudes $\phi_{cb}(k\sigma \ell)=
\text{Trace}\left\{|b\rangle \hat{c}_{k\sigma \ell}^{\dag}\langle
c|\hat{\rho}\right\}$. These $\phi_{cb}(k\sigma \ell)$ are themselves
determined by an equation of motion, see
Eq.~(7) of Ref.~\onlinecite{PedersenPRB2005a}.
In the first-order approach, all terms containing both $k$ and
$k'$ (correlations between two transitions) are
neglected and
Eq.~(10) of Ref.~\onlinecite{PedersenPRB2005a} has the solution
\begin{widetext}
\begin{equation}
\phi_{cb}(k\sigma \ell)(t)=
\frac{1}{\imai \hbar}\int_{-\infty}^t\d t'
\e^{\imai(E_b+E_k-E_c+\imai 0^+)(t-t')/\hbar}
\left(\sum_{b'}T_{cb'}(k)w_{b'b}(t')f_k
-\sum_{c'}w_{cc'}(t')T_{c'b}(k)(1-f_k)\right)
\label{EqPhiInt}
\end{equation}
Now we neglect the time dependence of $w_{b'b}(t')$ in the kernel
of the integral (the Markov limit) and set
$w_{b'b}(t')=w_{b'b}(t)$, which allows us to perform the integral.
Inserting this into the equation of motion for $w_{b'b}$, we
obtain
\begin{equation}\begin{split}
\imai \hbar \frac{\d }{\d t}w_{bb'}=&
(E_b-E_{b'})w_{bb'}\\
&+\sum_{a,k\sigma\ell}T_{ba}(k\sigma\ell)
\frac{\sum_{a'}w_{aa'}T^*_{b'a'}(k\sigma\ell)f_{\ell}(E_k)
-\sum_{b''}T^*_{b''a}(k\sigma\ell)w_{b''b'}(1-f_{\ell}(E_k))}
{E_k-E_{b'}+E_a-\imai 0^+}\\
&-\sum_{a,k\sigma\ell}
\frac{\sum_{a'}T_{ba'}(k\sigma\ell)w_{a'a}f_{\ell}(E_k)
-\sum_{b''}w_{bb''}T_{b''a}(k\sigma\ell)(1-f_{\ell}(E_k))}
{E_k-E_b+E_a+\imai 0^+}T^*_{b'a}(k\sigma\ell)\\
&+\sum_{c,k\sigma\ell}T^*_{cb}(k\sigma\ell)
\frac{\sum_{b''}T_{cb''}(k\sigma\ell)w_{b''b'}f_{\ell}(E_k)
-\sum_{c'}w_{cc'}T_{c'b'}(k\sigma\ell)(1-f_{\ell}(E_k))}
{E_k-E_c+E_{b'}+\imai 0^+}\\
&-\sum_{c,k\sigma\ell}
\frac{\sum_{b''}w_{bb''}T^*_{cb''}(k\sigma\ell)f_{\ell}(E_k)
-\sum_{c'}T^*_{c'b}(k\sigma\ell)w_{c'c}(1-f_{\ell}(E_k))}
{E_k-E_c+E_b-\imai 0^+}T_{cb'}(k\sigma\ell)\, .
\label{Eq1vn}
\end{split}\end{equation}
The current from the left lead into the system is given by
\begin{equation}
 J_L=-\frac{2}{\hbar}\Im\left\{\sum_{k\sigma,cb}
\frac{T^*_{cb}(k\sigma L)}{E_k-E_c+E_b+\imai 0^+}
\left(\sum_{b'}T_{cb'}(k\sigma L)w_{b'b}f_L(E_k)
-\sum_{c'}w_{cc'}T_{c'b}(k\sigma L)(1-f_L(E_k))\right)\right\}\, .
\label{EqJ1vn}
\end{equation}
\end{widetext}
If we restrict to diagonal elements $P_b=w_{bb}$, these equations
reduce to the standard master equation \cite{BeenakkerPRB1991}
formulated in a many particle basis
\cite{KinaretPRB1992,ChenPRB1994,PfannkuchePRL1995}.

The Redfield kinetics \cite{RedfieldIBM1957} has been recently
used to derive a similar set of equations \cite{HarbolaPRB2006}. We
can recover these equations, if we approximate
$w_{b'b}(t')=w_{b'b}(t)\e^{-\imai(E_{b'}-E_b)(t'-t)/\hbar}$ in
Eq.~(\ref{EqPhiInt}). In this case the integrations can be performed
as well, but we obtain slightly different denominators for
the nondiagonal elements in Eq.~(\ref{Eq1vn}).
While the approximation $w_{b'b}(t')=w_{b'b}(t)$ becomes exact in the
stationary state, which we consider here, the behavior
$w_{b'b}(t')\propto \e^{-\imai(E_{b'}-E_b)t'/\hbar}$ is suggested by
the linear term in the equation of motion.\footnote{For the
2vN-approach, the same ambiguity in closing the equations
arises. In this case, the analytical result for the single level
problem only matches the Green's function result, if the time dependence
is neglected, which is the approximation scheme used in
Ref.~\onlinecite{PedersenPRB2005a}.} At the moment, we have
no direct indication, which concept is more appropriate. However, we
did only find minor numerical differences and the qualitative features
are identical for all issues discussed in this article.
In particular, both approaches can yield negative probabilities
$w_{bb}$, a well-known problem of Redfield kinetics \cite{WeissBook1999}.

\section{Comparison of different approaches in the noninteracting limit}
\label{AppComparison}

Now we consider the double dot model without spin
in the noninteracting limit $U=0$, and set $E_{\alpha}=E_\beta=0$,
$\Gamma_L=\Gamma_R=\Gamma/2$ as well as $W\to\infty$.
In this case we can solve most approaches analytically, allowing for a
better understanding of the structure of the various approaches.

\subsection{Transmission formalism and 2vN approach}
As a bench mark, for the noninteracting case the current can be evaluated
exactly via the transmission formalism
(see e.g. Ref.~\onlinecite{DattaBook1995})
\begin{equation}
I=\frac{1}{2\pi \hbar}\int\d E \,
    T(E)[f_L(E)-f_R(E)]\label{EqCurrentT}
\end{equation}
with
\begin{equation}
T(E)=\frac{\Gamma^2\Omega^2}
{4[(E-\Omega)^2+\Gamma^2/16][(E+\Omega)^2+\Gamma^2/16]}
\end{equation}
where the wide band limit is applied. We obtained numerically the
same result from the 2vN approach for all parameters checked.

If $\Gamma\ll k_BT$, we may replace the peaks in the transmission
function by $\delta$-functions which provides us with
\begin{equation}
T(E)\approx
\left\{
\begin{array}{ll}
\frac{\pi\Gamma}{4}[\delta(E-\Omega)+\delta(E+\Omega)] &
\mbox{for}\, \Gamma\ll \Omega\\
\frac{2\pi\Omega^2\Gamma}{4\Omega^2+\Gamma^2/4}\, \delta(E) &
\mbox{for}\, \Omega \ll\Gamma \end{array}\right.\label{EqTapprox}
\end{equation}
The prefactor for $\Omega \ll\Gamma$ is chosen
such, that the integral over $E$ agrees with the full transmission
function for all $\Omega,\Gamma$.

\subsection{Master equation}
The master equation in the many-particle states
\cite{KinaretPRB1992,ChenPRB1994} can be derived by setting
$w_{bb'}=P_b\delta_{bb'}$ in the 1vN approach, resulting in the
current
\begin{equation}
I_{\text{master}}=\frac{\Gamma}{8\hbar}
\left[ f_L(\Omega)-f_R(\Omega)+f_L(-\Omega)-f_R(-\Omega)\right]
\end{equation}
which exactly equals the bench mark  (\ref{EqCurrentT}) in the limit
$\Gamma\ll k_BT,\Omega$, see Eq.~(\ref{EqTapprox}).

\subsection{Quantum rate equation}
Going beyond the master equation, correlations between different
states can be taken into account. The dephasing of these correlations
is frequently treated in a Lindblad form
\cite{WeissBook1999,LindbladCMP1976}, see, e.g.,
Refs.~\onlinecite{GurvitzPRB1996,StoofPRB1996,ElatariPLA2002,DongPRB2004}.
They can be derived in different ways and the
name ``quantum rate equation'' is frequently used.
Formulated in a basis of the localized states,
see e.g.  Ref.~\onlinecite{GurvitzPRB1996}
or Eq.~(36) of Ref.~\onlinecite{DongPRB2004}, we find the result
\begin{equation}
I_{\text{quantum rate}}=
\frac{\Gamma}{\hbar}\frac{\Omega^2}{\Gamma^2/4+4\Omega^2}(f_L(0)-f_R(0))\, ,
\end{equation}
which matches perfectly the bench mark result in the limit
$\Omega\ll \Gamma\ll k_BT$, see Eq.~(\ref{EqTapprox}). In addition, it is correct in the high-bias limit, $\mu_L,(-\mu_R)
\gg \Omega,\Gamma$, as proven by Gurvitz and
Prager.\cite{GurvitzPRB1996}

\subsection{1vN approach}
Finally the 1vN approach provides
\begin{equation}\begin{split}
I_{\text{1vN}}=&
\frac{1}{\hbar}
\frac{\Gamma\Omega^2}{\Gamma^2/4+4\Omega^2}
\frac{\left[f_L(\Omega)-f_R(\Omega)+f_L(-\Omega)-f_R(-\Omega)\right]}{2}\\
&+\frac{A}{8\hbar}\frac{\Omega\Gamma^2}{\Gamma^2/4+4\Omega^2} \label{EqI1vNanalytic}
\end{split}\end{equation}
where
\[
A=\frac{1}{\pi}\int^\infty_{-\infty}\d E \, [f_L(E)-f_R(E)]
{\cal P}\left\{
\frac{1}{E+\Omega}-\frac{1}{E-\Omega}\right\}
\]
is a small contribution from the real parts.
If the real parts are neglected ($A=0$), the result matches the bench
mark result as long as $k_BT>\Gamma$. This indicates that the
inclusion of the real parts may not be appropriate in the
noninteracting limit for a first order approach in the
  coupling $\Gamma$, see also the little peak at
$eV_\textrm{bias}= 2\Gamma$ for $U=0$ in
Fig.~\ref{FigCoulomb}. However, the comparison with the 2vN
approach (which does not display the spurious peak)
indicates that the real parts cover the  essential physics in the
interacting case. It is interesting to note, that the
Redfield kinetics provides exactly the same analytical result
(\ref{EqI1vNanalytic}), so that both approaches exhibit the same
problem in the noninteracting case.

\section{Comparison with Real-Time diagrammatic approach}\label{AppAnderson}
The 2vN approach allows for an analytic solution for the
single dot model with spin (the Anderson model) with
infinite Coulomb repulsion. We consider
a spin-degenerate level with
$E_{\up}=E_{\down}=E_d$.
Analogously to section III of Ref.~\onlinecite{PedersenPRB2005a},
we define
\[\begin{split}
B^{\ell}_{\sigma;0}(E)&=\sum_{k}\delta(E-E_{k})
T_{\ell}(k)\phi_{\sigma,0}(k\sigma \ell)\\
B_{\sigma;0}(E)&=B^{L}_{\sigma;0}(E)+B^{R}_{\sigma;0}(E)\\
\Gamma_{\ell}(E)&=2\pi\sum_{k}\delta(E-E_{k})T_{\ell}^2(k)\\
\Gamma(E)&=\Gamma_{L}(E)+\Gamma_{R}(E)
\end{split}\]
where we assume that the couplings to the contacts
$T_{\up 0}(k\up\ell)=T_{\down 0}(k\down\ell)=T_{\ell}(k)$ do not depend on
spin.
Then we obtain
\begin{multline}
\imai \hbar \frac{\d }{\d t}B^{\ell}_{\sigma;0}(E)=
(E_d-E+\Sigma(E)+\Sigma^f(E))B^{\ell}_{\sigma;0}(E)\\
+\frac{\Gamma_\ell(E)}{2\pi}\left[w_{0;0}f_\ell(E)-w_{\sigma;\sigma}
(1-f_\ell(E))\right]\\
-\frac{\Gamma_\ell(E)}{2\pi}\int\d E'
\frac{B_{\sigma,0}^*(E')+f_\ell(E)B_{\bar{\sigma},0}^*(E')}
{E-E'+\imai 0^+}
\label{EqBLDouble}
\end{multline}
where $\bar{\sigma}$ denotes the spin opposite to $\sigma$ and
\[\begin{split}
\Sigma(E_k)=&\sum_{k'\ell}\frac{T_{\ell'}(k')^2}
{E_k-E_{k'}+\imai  0^+}
\\
\Sigma^f(E_k)=&\sum_{k'\ell}\frac{f_{k'}T_{\ell'}(k')^2}{E_k-E_{k'}+\imai 0^+}
\end{split}\]
as well as
\begin{eqnarray}
\hbar \frac{\d }{\d t}w_{\sigma,\sigma}&=&-2\Im\left\{\int\d E
B_{\sigma,0}(E)\right\}\\
\hbar \frac{\d }{\d t}w_{0,0}&=&2\Im\left\{\int\d E
B_{\up,0}(E)+B_{\down,0}(E)\right\}\label{EqWup}
\end{eqnarray}
With the Ansatz $B_{\up;0}=B_{\down;0}=B(E)$ and
$w_{\up,\up}=w_{\down;\down}$ we have
the stationary solution
\begin{widetext}
\begin{multline*}
\frac{\Gamma_L(1+f_L(E))+\Gamma_R(1+f_R(E))}{2\pi}
\int\d E' \frac{B^*(E')}{E-E'+\imai 0^+}
=
(E_d-E+\Sigma(E)+\Sigma^f(E))B(E)\\
+\frac{w_{0,0}[\Gamma_L(E)f_L(E)+\Gamma_R(E)f_R(E)]}{2\pi}
-\frac{w_{\sigma,\sigma}
[\Gamma_L(E)(1-f_L(E))+\Gamma_R(E)(1-f_R(E))]}{2\pi}
\end{multline*}
As
\begin{equation}
\Im\{\Sigma(E)+\Sigma^f(E)\}
=-[\Gamma_L(E)(1+f_L(E))+\Gamma_R(E)(1+f_R(E))]/2
\label{EqImSigma}
\end{equation}
we find that there is a solution $B(E)$ which
is purely real (like in the spinless level case).
Inserting into
Eq.~(\ref{EqBLDouble}) gives
the stationary state
\begin{multline}
(E_d-E+\Sigma(E)+\Sigma^f(E))B^{L}_{\sigma;0}(E)=
\frac{\Gamma_L[1+f_L(E)](E_d-E+\Sigma(E)+\Sigma^f(E))B(E)}{\Gamma_L(E)(1+f_L(E))+\Gamma_R(E)(1+f_R(E))}\\
+\frac{\Gamma_L\Gamma_R(f_R(E)-f_L(E))}
{2\pi[\Gamma_L(E)(1+f_L(E))+\Gamma_R(E)(1+f_R(E))]}
\end{multline}
\end{widetext}
where we used $w_{0;0}+2w_{\sigma,\sigma}=1$.
With Eq.~(\ref{EqImSigma})
we have
\begin{equation}
\Im \left\{B^{L}_{\sigma;0}(E)\right\}=
\frac{\Gamma_L(E)\Gamma_R(E)(f_R(E)-f_L(E))}
{4\pi|E_d-E+\Sigma(E)+\Sigma^f(E)|^2}
\end{equation}
Finally the particle current (including both spin directions) is given
by:
\[
J_L=\frac{1}{\hbar}\int \d E\frac{\Gamma_L(E)\Gamma_R(E)(f_L(E)-f_R(E))}
{\pi|E_d-E+\Sigma(E)+\Sigma^f(E)|^2}
\]
This result fully agrees with the result from real-time perturbation
theory  in the so-called
resonant tunneling approximation
(containing a resummation of diagrams beyond second order perturbation
theory),  see Eq.~(4.61) of Ref.~\onlinecite{KonigDiss1998}.
This indicates that the 2vN approach contains an equivalent
set of higher than second  order tunneling processes
due to the self-consistency in  the equation of motion for $\phi_{cb}(k)$.

\bibliographystyle{apsrev}

\end{document}